\documentclass[12pt]{article}

\usepackage{color,epsfig,
amssymb
}
\begin{document}

\def\spose#1{\hbox to 0pt{#1\hss}}\def\lta{\mathrel{\spose{\lower 3pt\hbox
{$\mathchar"218$}}\raise 2.0pt\hbox{$\mathchar"13C$}}}  \def\gta{\mathrel
{\spose{\lower 3pt\hbox{$\mathchar"218$}}\raise 2.0pt\hbox{$\mathchar"13E$}}}

\def\thf{\baselineskip=\normalbaselineskip\multiply\baselineskip
by 11\divide\baselineskip by 10}
\def\fff{\baselineskip=\normalbaselineskip} 

\thf

\definecolor{turq}{rgb}{0,0,0}
\definecolor{violet}{rgb}{0.4,0,0.6}
\def\Inf{ {\color{violet}{I}} }
\def\calM{{\color{violet} {\cal M}} }
\def\calT{{\color{violet} {\cal T}} }
\def\mmu{{\color{violet} {\cal A}} }
\def\Tee{{\color{violet} {T}} }

\def\mm{{\color{red}m}} 
\def\ee{{\color{red}e}}

\def\ff{{\color{red}f}}
\def\alp{{\color{black}\sigma}} 

\def\kkappa{{\color{red}\kappa}}
\def\varth{{\color{red}\vartheta}} 
\def\vepsilon{{\color{red}\varepsilon}}

\def\aa{{\color{blue}a}}
\def\RR{{\color{blue}R}}
\def\tt{{\color{blue}t}}  
\def\ttau{{\color{blue}\tau}}

\def\PP{{\color{red} {\cal P}} }
\def\Nn{{\color{red} {\mathfrak N}} }
\def\NN{{\color{red} {N} }}
\def\mur{{\color{red} {\mu}}} 
\def\rr{{\color{red} {u} }}
\def\ss{{\color{red} {s} }}

\def\Qq{{\color{red} {\mathfrak q}} }

\def\be{\begin{equation}}
\def\fe{\end{equation}}

\bigskip

\centerline{\Large \bf Hominid evolution: genetics versus memetics.}
\bigskip

\centerline{\bf Brandon Carter}

\medskip

\centerline{CNRS, LuTh, Observatoire de Paris,}

\centerline{92195, Meudon, France.}

\bigskip
\centerline {(Definitive version for Int. J. Astrobiology.)}
\medskip
\centerline{\bf  June, 2011.}

\vskip 1 cm

\noindent

{\bf Abstract:} 

The last few million years on planet Earth have witnessed two
remarkable phases of hominid development, starting with  a phase 
of biological evolution characterised by rather rapid increase of 
the size of the brain. This has been followed by a phase of even more 
rapid technological evolution and concomitant expansion of the size 
of the population, that began when our own particular `sapiens'  
species emerged, just a few hundred thousand years ago. The present 
investigation exploits the  analogy between the neo-Darwinian genetic 
evolution mechanism governing the first phase, and  the memetic 
evolution mechanism governing the second phase. From the outset 
of the latter until very recently -- about the year 2000 -- the growth 
of the global population $\NN$ was roughly governed  by an equation of the 
form ${\rm d}\NN/\NN\,{\rm d} t= \NN/\Tee_{\! *}$, in which  $\Tee_{\! *}$
 is a coefficient introduced (in 1960) 
by von Foerster, who evaluated it empirically as about two hundred 
thousand million years. It is shown here how the value of this
hitherto mysterious timescale governing the memetic phase is
explicable in terms of what happenned in the preceding genetic phase. 
The outcome is that the order of magnitude of the Foerster timescale 
can be accounted for as the product of the relevant (human) generation 
timescale, about 20 years, with the number of bits of information in
the genome, of the order of ten thousand million. 
Whereas the origin of our `homo' genus may well have 
involved an evolutionary hard step, it transpires that the emergence
of our particular `sapiens' species was rather an automatic process.

\vskip 2 cm


\vfill\eject 
\noindent
{\bf 1. Introduction}
\medskip

Deeper understanding of the evolution of life  on planet Earth is
of interest not only in its own right but also for the light
it can throw on what can be expected in the extra solar planetary 
systems that are being discovered at an increasing rate. Reciprocally, 
consideration of what may happen elsewhere can help provide a deeper 
understanding of what has already happened -- and what is likely
to happen -- in our own terrestrial case. This applies particularly 
to anthropic selection effects, as exemplified by the significance, 
with respect to the question of the hard-steps 
(Carter 1983, 2008; Watson 2008; McCabe 2010) in our evolution, 
of the empirically observed coincidence that the age of the Earth is 
comparable with the expected total (past and future) life of the Sun.

The present article is concerned with two other -- independent but
clearly related  -- empirically observed coincidences that seem 
likely to be relevant to the question of what was the last of these 
hard-steps. An obviously plausible candidate for the status of the 
last hard-step is the process that occurred when our ancestral line 
branched off from that of the chimpanzees round about six million 
years ago. This step is the latest of the 39 bifurcations listed by 
Dawkins (2004) (whose introductory presentation of the anthropic 
principle did not address the terrestrial hard-step question, but 
got sidetracked into far fetched cosmological speculations).
However it is also conceivable that this evolutionary step
-- the onset of the first main phase of hominid evolution --
was not extremely hard (in the sense of being highly improbable
within the limited time available) but that the last hard-step
occured at a more recent occasion, for which the most obvious 
candidate is the onset of the second main phase of hominid 
evolution, as marked by the emergence of our own species only a 
few hundred thousand years ago.

On the basis of the meagre palaeontological evidence available, the 
salient feature of the first phase was systematic growth of cranial 
(and presumably corresponding intellectual) capacity, which proceeded 
at a modest rate in the genera australopithecus and paranthropus, and 
then became remarkably rapid (Falk,1998; Holloway 2001) after our own 
genus `homo' had branched off -- at what is a plausible alternative  
candidate for hard-step status -- a couple of million years ago. 

The second phase started relatively recently, when our own particular 
species, `homo sapiens' finally emerged, just a few hundred thousand 
years ago. Instead of the genetic evolution that characterised the 
previous phase, this second phase -- which has lasted until about now --  
has been characterised by technological evolution and concomitant 
population expansion to fill the increasing range of newly created 
ecological niches. Such evolution is describable as memetic, because 
the technological know-how on which it depends is analysable in terms 
of memes,  meaning replicable cultural information units, a fruitful 
concept originally introduced by Dawkins (1976) who drew attention
to the analogy between memetic evolution and ordinary genetic evolution 
as described by neo-Darwinian theory. 

This second phase of hominid evolution has recently culminated
in a global `high tech' civilisation characterised by the first
of the apparently coincidences referred to above, which is that
the human population $\NN$ has reached a value of the same order 
of magnitude as the number $\Inf$ of bits of information in the genome,
which for ordinary terrestrial (DNA programmed) animals is of the order
of a several Giga (using the  unambigous Greek based term `Giga'  for 
what the French call a ``milliard'' and what the Americans call a 
``billion'', namely a thousand million, as distinct from the 
original meaning of the word billion which was a million million).

The consideration that this large number coincidence, 
{\be \NN\approx \Inf \, , \label{firstcoin} \fe}
 is not something that held at other times in the past, but something 
valid only at a particular period, namely or own, suggests that it 
should be accounted for as an anthropic selection effect.

The status of the related coincidence referred to above is rather 
different, as it relates the same number $\Inf$, not to the 
present value of a variable, but to a constant, namely the coefficient 
$\Tee_{\!\star}$ in the hyperbolic growth formula,
{\be \dot \NN=\NN^2/\Tee_{\!\star} \, ,\label{quadgrowth}\fe}
which provides a remarkably good order of magnitude estimate
for the rate of growth $\dot \NN$ of the human species ever since
our emergence a few hundred thousand years ago, when $\dot \NN$
would have been only a few people per year right up to the
present time when $\dot \NN$ is not far from a hundred million
people per year. The applicability of this extremely
simple formula (\ref{quadgrowth}) for a roughly  constant
value $\Tee_\star\approx 2\times 10^{11}$ was recognised
about half a century ago by von Foerster (1960),
and its conceivable extraterrestrial relevance was pointed out 
soon afterwards by von Hoerner (1975).
The fact that $\dot\NN$ is proportional to $\NN^2$ (rather than 
linearly proportional to $\NN$ as in the ordinary exponential case) 
has more recently been shown to be loosely accountable for in 
terms of the essentally memetic nature of the process whereby 
technical progress provided an accelerating increase in the 
population carrying capacity of the planet (Kremer 1993; 
Koratayev 2005). However the reason why the Foerster coefficient
$\Tee_{\!\star}$ has its particular value has remained mysterious.
The motive of the present work is to show that a clue to this
enigma may be obtainable from the second of the empirically 
observed coincidences referred to above, which is that the
Foerster coefficient is given in terms of the human generation
timescale $\ttau_{\!g}$ by the order of magnitude relation
{\be \Tee_{\!\star} \approx \Inf\, \ttau_{\!g}\, .\label{secondcoin}\fe}

This second coincidence  - which seems to be a permanent feature of the
human population - provides an immediate clue to the 
explanation of the first coincidence (\ref{firstcoin}), since
it is interpretable as meaning that the decreasing  population growth
timescale $\ttau=\NN/\dot\NN$ has now got down to the order of magnitude
its Malthusian minimum, $\ttau\approx\ttau_{\!g}\, ,$ which evidently means
that the Foerster phase characterised by (\ref{quadgrowth}) can continue
no longer. The concluding section of this article will discuss the
implication of our presence at this critical moment, which will be
anthropically explicable if the population peaks and then declines
in the not so distant future.

The main part of this article will investigate the way the second 
coincidence (\ref{secondcoin}) may itself be accounted for by taking 
the analogy between genetics and memetics seriously, on the basis 
of the presumption that the relevant selection pressure -- 
favouring increasing mental capability during the first phase, 
and increasing technological capability during the second phase 
-- would have been sufficiently high to bring about progress at
a rate that would on both cases  have been proportional to the size 
$\NN$ of the relevant interbreeding  population, as explained 
on the basis of the simplified model of neo-Darwinian evolution 
that is outlined in Section 6.

\bigskip
\noindent
{\bf 2. Extrapolation by the seductive Verhulst model}
\medskip

\begin{figure}
\centering
\epsfig{figure=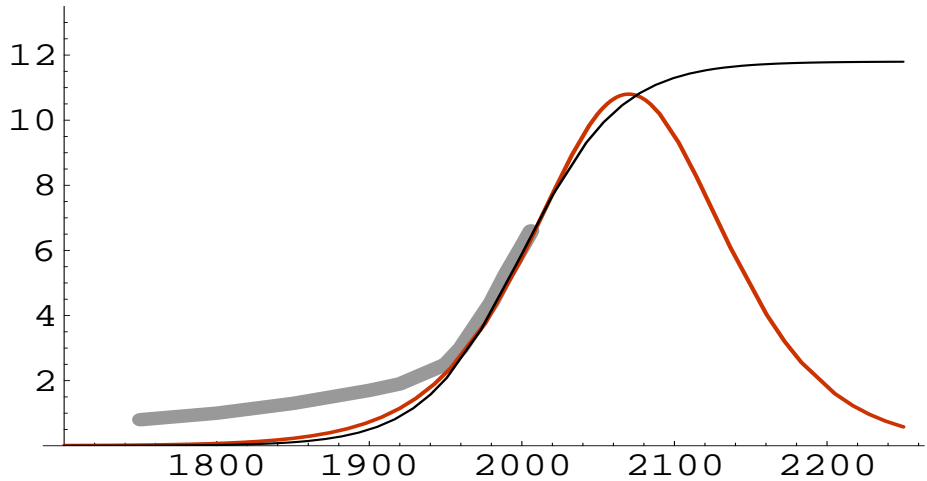, height=6.4 cm, width=12.8 cm}
\caption
{Global population $\NN$, in units of  $10^9$, plotted against date.
Thick pale curve shows known U.N. data from
1750 A.D. to  2000 A.D.  Thin dark curve shows a commonly considered
future extrapolation by Verhulst type  logistic model 
(\ref{4}), for $\tau\simeq 32$ yr, fitted  to the latest observations, 
with inflection about 2000 A.D. Medium pale  curve shows an alternative 
extrapolation by Hubbert type logistic derivative model (\ref{9}), 
for $\tau\simeq42$ yr, with peak time $\tt_p$ about 2070 A.D., as 
conjectured by Bourgeois-Pichat. Neither model fits the data for the 
past.
}
\label{extrapolate}
\end{figure}

It is to be recalled that the theory of demographic growth has a 
history dating back to the work of Malthus, who introduced the 
simplest and still most widely used kind of model for this purpose, 
namely that of the exponential type, as obtained from an evolution 
equation of the form 
{\be \dot\NN=\NN/\ttau\, ,\label{expev}\fe} 
with a fixed e-folding timescale $\ttau$ so that the solution will 
take the form
{\be \NN=\NN_{\!_0}\,{\rm exp}\{\tt/\ttau\}\, ,\label{0}\fe}
where $\NN_{\!_0}$ is the population at some chosen time origin
when $\tt=0\, .$ Having pointed out that it is biologically possible 
in principle for the population to grow exponentially with a timescale
$\ttau$ that could  be as short as the reproductive generation 
timescale $\ttau_{\!g}\, ,$ -- about 20 years in the human case -- 
Malthus drew the conclusion that the relatively slow growth observed 
in practice was attributable to (typically unpleasant) causes like war, 
disease, and particularly famine, as an ineluctable consequence of the  
limited availability of food and other necessities. The corollary that, 
as well as luck, the survivors were likely to be characterised by 
superior aptitudes that would often be hereditary was the basis of 
Darwin's theory of evolution by natural selection.

The recipe for perpetual exponential growth is still commonly
sought as an ideal  ``holy grail'' by economists, despite the 
recognition  of its unsustainability in the long run, by Malthus.
The first and simplest ``sigmoid'' model allowing for the limited 
availability of renewable resources was introduced by his follower, 
Verhulst, in the middle of the nineteenth century, but
it took another century before attention began to be given to the
need to take analogous account of  the limited availability of
non-renewable resources, for which a corresponding ``peaked''
model was introduced by Hubbert.

The simple ecological model due to Verhulst is based on
an evolution equation of what is known as the logistic form,
{\be \dot \NN/\NN=(1-\NN/\NN_{\!_\infty})/\ttau\, , \label{3}\fe}
for some fixed saturation value $\NN_{\!_\infty}\, ,$
interpretable as the maximum environmentally sustainable value of
$\NN$.  (For example, if one allows about an acre for a family of
four, the entire land surface of the world gives $\NN_{\!_\infty}
\approx 10^{11}$.) The solution of the Verhulst
equation (\ref{3}) will be symmetric with respect to a time $\tt_s$ 
in terms of which it takes the well known logistic form
{\be \NN=\frac{_1}{^2}{\NN_{\!_\infty}}\left(1+{\rm tanh}\left\{
({\tt-\tt_s})/{2\ttau}\right\}\right) \, ,\label{4}\fe}
which shows how the upper bound $\NN_{\!_\infty}$  will be
asymptotically approached from below. The smoothly controlled
Verhul{\color{turq}st} model was originally intended for application, 
not to the world as a whole, but just to the newly established
kingdom of Belgium, for which it was remarkably successful.
However -- as an example of the more unpleasant alternatives about
which Malthus had warned -- in Ireland about the same time,
exponential growth was terminated, not by smooth convergence to
a plateau level, but catastrophically (Catton 1980) by a  famine.

The Verhulst model is still popular for extrapolation 
of present data to construct demographic predictions of the kind 
commonly provided by U.N. and other international organisations as 
shown in Figure \ref{extrapolate}. However as the twentieth century 
advanced, people became increasingly aware that as well as the 
saturation on renewable resources, another important limitation was 
the exhaustion of non-renewable resources. 
\bigskip

\noindent
{\bf 3. Population time, anthropic measure, and the Hubbert model}
\medskip

To take account of the finiteness of non-renewable resources,
it is useful to think, not so much in terms of ordinary 
time $\tt$, but rather in terms of population time  $\calT \, $ 
meaning the total of the times lived by all the members of 
the population, as given by the condition that its increment 
${\rm d}\calT$ during a small time interval ${\rm d}
{\color{turq}t}$ is given in terms of the relevant population 
size $\NN$ by  ${\rm d}\calT= \NN\, {\rm d}\tt\, ,$ so that 
its rate of increase is
{\be \dot\calT=\NN\, .\label{poptimerate}\fe}
This concept of the population time, as defined by 
(\ref{poptimerate}), was introduced by Wells (2009) in a 
discussion of the implications for future demographic prospects
of the anthropic principle.

{\color{turq} The idea underlying the anthropic principle 
(Carter 1983; Leslie 1996; Bostrom 2002) is that to interpret what 
one observes when one emerges in the universe one needs some idea 
of what might or might not be expected as far as one's personal 
identity and situation in space and time are concerned. For example 
one ought to be more surprised to find oneself to be a prince than 
to find oneself in the (less exclusive) category of peasants. As 
originally formulated for application in a cosmological context, 
with respect to conceivable extraterrestrial observers, the (weak) 
anthropic principle merely postulated vaguely that a priori 
probability should be democratically distributed among comparable 
observers. For many applications that was enough, but for other 
purposes one needs to clarify the question of what counts as an 
`observer', and of how one should distribute a priori probability 
among observers that are not `comparable'.  It might be consensually 
accepted that a prince should be considered to be comparable with 
a peasant, but it is clear that neither is comparable with a cat, 
and indeed it is not entirely clear whether a cat should be counted 
as an `observer' at all.

To cope with such unresolved issues, it is convenient to introduce 
an appropriately adjustable parameter that I shall refer to as 
the anthropic quotient. Thus the general purpose version that 
I would now advocate for the anthropic principle postulates that} 
the {\it a priori} probability per unit time of finding oneself to be 
a member of a particular population is proportional to the number of 
individuals in that population, multiplied by an anthropic quotient 
$\bar\Qq$ say, that will depend on the kind of population involved. 
The relevant probability will thereby be proportional to the 
corresponding anthropic measure, $\mmu$ say, for which the 
differential increment will be given by 
${\rm d}\mmu= \bar\Qq\, {\rm d}\calT\, , $ so that
its rate of increase will be $\dot\mmu=\bar\Qq \NN\, .$

There is no loss of generality in fixing the normalisation
of the anthropic measure $\mmu$ by the obviously suitable
convention that $\bar\Qq$ be set to unity, $\bar\Qq=1 \, ,$
in the ordinary human case, with which Wells was concerned. The 
appropriate value of the anthropic quotient might exceed unity
for conceivable superhuman extraterrestrials, but it should
presumably be smaller, $\bar\Qq<1$ for our hominid ancestors
such as homo erectus, and even more so, $\bar\Qq\ll 1\, ,$
for our subsisting terrestrial anthropoid relations, such as
chimpanzees (not to mention cats).

As well as being proportional to anthropic probability, in the 
case under consideration in the present section, that of the 
modern  humans with $\bar\Qq=1$, the elapsed population time 
$\calT$ is also proportional to the amount of non-renewable
resources that have been consumed, with a proportionality
coefficient that depends on the nature of the resource and
on the level of development of the population. In the particular 
context of oil extraction, it was recognised by Hubbert 
(of the Shell company) and other engineers involved that models
of the logistic  kind described above could be applicable
to accumulated consumption. This implies that during a period
characterised by a fixed per capita consumption rate such models 
would be applicable not to $N$ as in the Verhul{\color{turq}st} 
case but to $\calT \, ,$ which would be subject, in such 
circumstances, to an evolution equation of the form
{\be \dot \calT/\calT=(1-\calT/\Tee_{_\infty})/\ttau\, , \label{7}\fe}
for some constant $\Tee_{_\infty}$ proportional to the total
amount of population time that can be lived before exhaustion
the ressources of the resources in question.

As the analogue of (\ref{4}), the solution of this equation (\ref{7})
will have the  logistic form
{\be \calT=\frac{_1}{^2}\Tee_\infty\left(1+{\rm tanh}\left\{({\tt-\tt_p})
/2\ttau\right\}\right)
\, ,\label{8}\fe}
in which $\tt_p$ is  a constant of integration that is interpretable as
the time at which the corresponding total population
{\be \NN=\Tee_\infty/4\ttau\,{\rm cosh}^2
\left\{({\tt-\tt_p})/2\tau\right\}
\, ,\label{9}\fe}
reaches its peak value, namely $\NN_{\!p}=\Tee_\infty/4\ttau$,
after which it undergoes a smooth transition towards an
asymptotic state of exponential decay.

A scenario characterised by a demographic transition towards exponential 
decay in a manner similar to that of the Hubbert model has also been 
predicted  on the basis of a rather different mechanism by  
Bourgeois-Pichat (1988). In scenarios of the traditional kind, 
private investment in child raising is motivated as an 
insurance against destitution in old age, and is affordable by 
the majority because education is either neglected or -- in more 
civilised societies -- provided at public expense. On the contrary, in 
a  Bourgeois-Pichat type scenario, the need for such (private) 
insurance is reduced by public support for the elderly (who 
will be relatively numerous and politically preponderant in 
a numerically declining population) while on the other hand the 
maintenance of a high standard of living requires a level of education 
that makes child raising unaffordable -- even with public assistance --
for many people. A population implosion of the kind that ensues 
in such circumstances has already been observed on a local scale
in some of the most ``developed'' countries, starting with (West) 
Germany, but the supposition by Bourgeois-Pichat that this will also 
happen in the rest of the world has not yet been confirmed.

Although -- for such diverse reasons -- a Hubbert type model may 
 conceivably provide an adequate description of the future, it can be 
seen from Figure \ref{extrapolate} that it fails completely for 
describing the global population in the past.  As -- like a 
Verhul{\color{turq}st} model -- its initial comportment is that of 
exponential growth, a Huppert type model vastly underestimates the 
population that was actually present in the distant past. One of the 
reasons for this failure is of course that the assumption of a constant 
per capita consumption rate did not apply: on the contrary, before the 
industrial revolution the global consumption of non-renewable 
resources was negligible.

As well as being incompatible (albeit for other reasons) with what is 
known about the past, the Verhul{\color{turq}st} type extrapolation is 
much less credible, on anthropic grounds, than its Hubbert analogue 
 as a description of the long term future. In order for a model to 
be plausible, our situation within it should not be too unlikely a 
priori. According to the anthropic principle a  general (necessary, 
but not necesssarily sufficient) requirement for plausibility of 
a demographic model will be a finitude requirement to the effect 
that the total human  population time $\cal T$ of the future should 
not greatly exceed that of the past and vice versa. This condition 
is evidently satisfied by a scenario of the kind foreseen by
 Bourgeois-Pichat (1988) as represented by the Hubbert type model 
in Figure \ref{extrapolate}, for which it can be seen that the 
population time $\calT\, ,$ (representing the area under the curve) 
and hence the corresponding anthropic measure, converge to the finite 
limit $\Tee_\infty$, which is only about twice the value attained 
already. However for the Verhul{\color{turq}st} type model the 
population time will diverge linearly according to a formula of the 
asymptotic form $\calT  \sim \NN_{\!_\infty}\tt\, .$  Thus  according  
to what is known as the ``doomsday argument'' (Leslie 1996; Bostrom 
2002; Wells 2007) such a Verhul{\color{turq}st} type extrapolation 
can be credible only locally, subject to the proviso of being 
truncated by a ``doomsday'' cut-off in the not too distant future.

For purposes of demographic prognostication, the basic anthropic
finitude condition to be satisfied is that the total human 
population time $\cal T$ of the future should not greatly
exceed that of the past. In order to apply this principle we 
therefore need a reliable estimate of how much human population 
time $\calT$ has elapsed so far. Such an estimate can not be 
provided by foregoing logistic models, but is obtainable from the
Foerster model as recapitulated in the next section.

\bigskip
\noindent
{\bf 4. The Foerster model: a good fit for the past.}
\medskip

It was reasonable for economists and social scientists such as
Verhul{\color{turq}st}, and other early  followers of Malthus, to 
seek timescales  of the order of a human lifetime, or at most of 
the duration of human history, for the formulation of their 
demographic models. When simple models involving a single
such timescale $\ttau$ were found to be inadequate, they 
resorted (Cook 1962) to elaborate multi-timescale models 
with too many adjustable parameters to be of much help for 
prediction. It was hard to see that the available demographic 
data were after all describable very well in terms just of a
single timescale, $\Tee_\star$, because the required value
is literally astronomical. It is therefor unsurprising that 
the first to have recognised it should have been not an economist, 
or even an ecologist, but a physicist, Heinz von Foerster 
(1911-2002) from Vienna, who noticed at last (Foerster 1960)
that the available demographic data could be fitted rather well by 
a formula of the simple hyperbolic form
{\be \NN= \frac{\Tee_{\!\star}}{\tt_d-\tt}\, , \label{12}\fe}
which is exactly what is obtained from the evolution law
(\ref{quadgrowth}) derived above, subject to the specification of a 
divergence time $\tt_d$ that arises as a constant of integration.

The validity of this formula  -- as a fairly good 
approximation with a roughly constant value of $\Tee_{\!\star}$
all the way from palaeolithic to modern times --
did not become widely known until relatively recently, and is 
something I observed independently, before finding out that it 
had already been pointed out by von Hoerner (1975),  
and  before that by von  Foerster (1960), who estimated
that the remaining time  before the singularity was then barely
70 years. More than half that time has since been used up,
but the remarkable -- and rather alarming -- fact is that significant 
deviation from the Foerster formula has not yet become clearly
observable.

\begin{figure}
\centering
\epsfig{figure=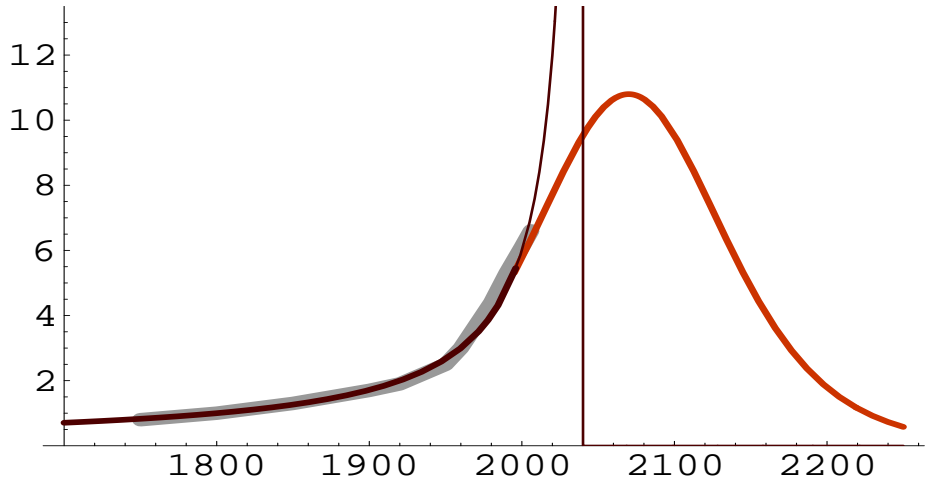, height=6.4 cm, width=12.8 cm
}\caption
{Population $\NN$, in units of  $10^9$, plotted against date.
Thick pale  curve shows U.N.  data for period before 2000 
A.D.  Medium pale curve, after that date,  shows Bougeois-Pichat 
type extrapolation given by a Hubbert model, as in  Figure 
\ref{extrapolate}. Black curve shows Foerster model (\ref{12}) 
for $\Tee_{\!\star}\simeq 24\times 10^{10}$ yr, with divergence 
date $\tt_d \simeq 2040\, , $ which agrees well with data before  
2000 A.D.
}
\label{FigFoerst}
\end{figure}

Theoretical explanations of  the acceleration from an initially 
slow start (what has been referred to by Renfrew (2008) as the ``sapient 
paradox'') and more particularly of the quadratic form, 
$\dot\NN\propto \NN^2$ of the relevant growth law  (\ref{quadgrowth}), 
have been proposed by Kremer (1993)  and Koratayev (2005) in terms of 
theories of technological development along lines of the kind to be 
sketched in Section 7, but with the coefficient $\Tee_{\!\star}$ 
introduced just as an adjustable parameter to be fixed empirically by 
matching to what is observed.
Substitution of the value  $\Inf\approx 10^{10}$  with 
$\tau_{\!g}\simeq 20$ years in the formula (\ref{secondcoin}) gives 
a result that is in good agreement with the more precise value
{\be \Tee_{\!\star}\simeq 240\  {\rm Gyr} \, . \label{11}\fe}  
that I obtain, as shown Figure \ref{FigFoerst}, by matching the formula 
(\ref{12}) to the official statistics  up to about 
2000 A.D. (U.N. Population Division 1999) with the correspondingly 
adjusted value of the constant of integration -- namely the divergence 
date -- given by {\be \tt_d\simeq 2040\ {\rm A.D.} \label{13}\fe}

It is to be remarked that, on the basis of fine tuning to the 
demographic statistics of their own time in the short run, von 
Foerster (1960) and von Hoerner (1974) originally  suggested a 
``doomsday'' time that was even nearer, $ \tt_d\simeq 2025 $ A.D., 
in conjunction with a fixed timescale that was correspondingly 
reduced, $\Tee_{\!\star}\simeq 200$ Gyr. However the rather longer 
fixed time scale (\ref{11}) and the rather later divergence 
time (\ref{13}) seem to give a better match in the long run, 
not just for more recent years, but also for the more distant 
past, through mediaeval times. For even earlier (classical, bronze 
age,  neolithic, and palaeolithic) times (Cook 1962; Biraben 1983)
the uncertainties are anyway so large that the differences between 
such alternative adjustments are not statistically significant. 

\bigskip
\noindent
{\bf 5. The measure of the Foerster phase.}
\medskip

According to (\ref{12}) and (\ref{11}) the size of the global 
population at the time, $\tt_1$ say, when our own species first 
emerged, a few hundred thousand years ago, would have been given 
roughly by $\NN_1\approx 10^6\, $ an order of magnitude that is 
consistent (Kremer 1993) with the the little that is known 
(Biraben 2003) about that epoque. Much of that total would not have 
been direcly ancestral to ourselves, but would  have included various 
`erectus' and Neanderthal  populations, as well as many fragmented 
`sapiens' groups that subsequently died out without leaving any 
descendants (which is why the ``effective'' ancestral population size 
obtained (Wade 2007) from the analysis of the modern human genome is 
very much smaller, only a few tens of thousands). In terms of this 
initial value $\NN_1$ the subsequent measure, attributable almost 
entirely to our own species,  will be expressible as 
{\be \calT=\Tee_{\!\star}\, {\rm ln}\{\NN/\NN_1\}\, ,\label{18}\fe}
which means that $\NN/\NN_1$ grows as an exponential function,
not of ordinary time but of the population time ratio 
$\calT/\Tee_{\!\star}\, .$

Up to the present time  the expansion factor $\NN/\NN_1$ is about  
$10^4\simeq {\rm e}^{10}$, so the ordinary (decimal) logarithm
 of the ratio is ${\rm log}\{\NN/\NN_1\}\simeq 4\, ,$ and the
corresponding natural logarithm is 
{\be  {\rm ln}\{\NN/\NN_1\}\simeq 10\, ,\label{tenfactor}\fe} 
so it follows from (\ref{11}) and (\ref{18}) that the  
population time measure of the whole of our `sapiens' species 
until now -- as required for the application (Wells 2007) of 
the ``doomsday argument'' -- is given roughly by 
$\calT\approx 2.4\times 10^{12}$ human years. 

\begin{figure}
\centering
\epsfig{figure=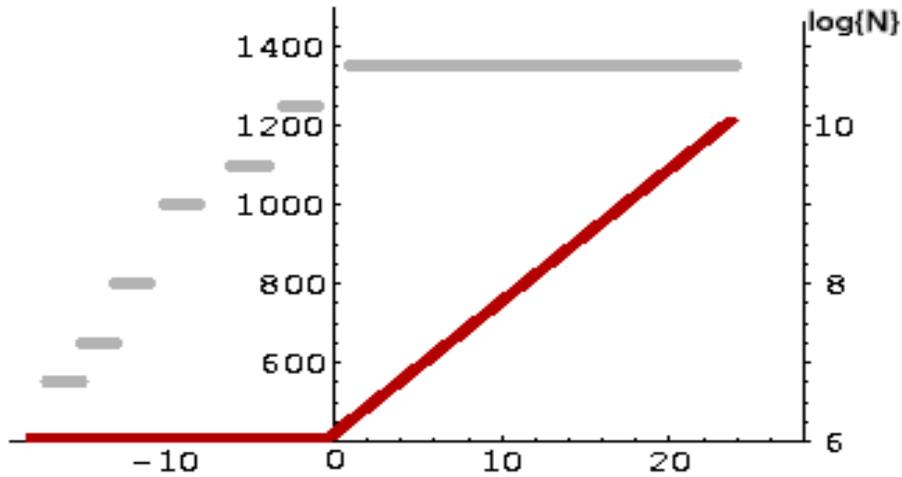, height=6.4 cm, width=12.8 cm
}\caption
{Evolution of the genus `homo' in terms of population 
time $\calT \, ,$ as measured in Giga centuries, from an origin at the 
emergence of our `sapiens' species, when the Foerster phase began. 
Schematic (order of magnitude)  plot of logarithmic population size, 
${\rm log}\{\NN\}\, ,$ is shown by thick firm line.
 Brain size in cm$^3$ is indicated, by thick pale horizontal segments, 
for successive representatives of the genus homo, namely the species 
robustus, habilis, ergaster, Java (archaic erectus), erectus, 
Heidelberg, and finally sapiens.
}
\label{poplot}
\end{figure}

The linear growth of ${\rm log}\{\NN\}$ from about 6 to about 10
is plotted in Figure \ref{poplot} against the hominid population 
time $\calT\, ,$ as roughly measured from an origin at the (calender) 
time $\tt_{\!_1}$ when our `homo sapiens' species first emerged a 
couple of hundred thousand year ago or so ago.

Before then, since the emergence of the `homo' genus, about a
 couple of million years earlier, it is thought (Biraben 2003)
that the total population would have fluctuated about a roughly 
fixed value, of the order a million, so  the population time 
$\calT$ would have been roughly proportional to ordinary time $\tt$:
a hundred thousand years of $\tt$ would have represented
about a Giga century of $\calT$. It is instructive to measure
human population time in centuries as that gives a rough indication of
the number of people who lived complete lives. (The number of people who
lived at all would of course be considerably higher, as life expectation
was low, more comparable with a human generation timescale of only about
20 years, so  a single human century would have typically represented
as many as five distinct ``souls''. However most of those five would 
have died before emerging from childhood, while only about one of them
would have survived long enough to lead a full adult life.)

Although the total population size for the the genus `homo' did not 
change much during the period ${\calT}\lta 0$ (before the emergence of 
our `sapiens' species) the nature of the population underwent rapid 
evolution {\color{turq} (driven, presumably, by intraspecific 
competition) of which the salient feature was the remarkably rapid
increase in brain size, as indicated by cranial capacity.} 

The growth (Falk 1998;  Holloway 2001)  of brain size, as measured 
in cm$^3$ is plotted in Figure \ref{poplot} for successive species, 
starting with homo robustus (which coexisted with survivors of the 
prevously dominant hominid genus, namely australopithecus). 
The figure shows that after
continuing at a rather high and steady rate so long as $\calT$ was 
negative, this brain growth came to a rather abrupt halt, {\color{turq} 
having reached a value of about 1400 cm$^3$, not long after} the time 
when systematic population growth got under way. During the period  
${\calT}\gta 0 \, ,$ namely that of the Foerster phase, brain growth was 
superseded by the population growth that took off -- at a comparable 
rate --  after the emergence of the two latest species of homo, 
namely our own `sapiens' species and the less long lived Neanderthal 
species (which  is not indicated separately in Figure \ref{poplot} 
because it was effectively  overlapped, both in brain size and 
duration, by `homo sapiens').

As well as the remarkable albeit approximate synchronisation of the 
end of brain growth with the acceleration (if not quite the beginning) 
of population  growth, Figure \ref{poplot} also features a rather 
striking numerical coincidence, which is that with respect to 
population time (not ordinary time) the brain growth is characterised 
by a timescale, $\Tee$ say, that is of the same order of magnitude 
as the timescale $\Tee_{\!\star}$ characterising population growth
according to (\ref{quadgrowth}). 

It be shown in 
the following sections how these features provide a clue for 
understanding  the previously remarked coincidence (\ref{secondcoin}) 
relating $\Tee_{\!\star}$ to the genome information content 
$\Inf \, . $  The essential link is that the latter is obviously 
involved in the genetic evolution process responsible for the brain 
growth caracterised by $\Tee\, .$

\bigskip
\noindent
{\bf 6. Simple  neo-Darwinian modelisation}
\medskip


Although satisfactory for the description of large bodies, classical 
physics as developed before the twentieth century was inadequate for 
the description of smaller systems, which need allowance for atomic 
substructure and the use of quantum mechanics. In an analogous way, 
classical Darwinian theory -- treating evolution as a continuous 
process -- is adequate only for very large populations.  A less naively 
simple description is needed for small and medium sized populations, 
meaning those in which
{\be \NN\lta \NN_{\!\rr}\, , \label{smallpop}\fe}
where $\NN_{\!\rr}$ is the replication reliability number, defined as 
the number of successive generations over which  one would expect 
the genetic information at a particular locus to be reliably
copiable, as given in terms of the corresponding
  mutation rate $\rr$  by
{\be \NN=1/\rr\, .\fe} 
 
The discrete nature of genetic information, was first pointed 
out by Mendel in Darwin's time, but it was was not until after 
Morgan's observational discovery of the mutation process that 
the neo-Darwinian theory needed to allow for the finiteness of the 
mutation rate (Maynard Smith 1989) was developed  by pioneers 
such as Fisher and Wright, in terms of general principles that 
would presumably be valid for other life systems, which might 
conceivably record the relevant hereditary information in some 
alternative form instead of DNA, whose use in the terrestrial case 
was subsequently discovered by Watson and Crick. The terrestrial
system uses a 4 letter code so that the information content at each
base position on the chain is only 2 or 3 bits depending on whether
the DNA chain is duplicated, in the manner that is normal for the
eukaryotic cells of multicellular plants and animals. This means
that the total number, ${\calM}$ say, of base positions is of the
same order of magnitude as the total genome information content,
$\Inf\approx {\calM}\, ,$ a relation that might be expected to hold
also for conceivable extraterrestrial systems using some alternative 
to DNA. 

\begin{figure}
\centering
\epsfig{figure=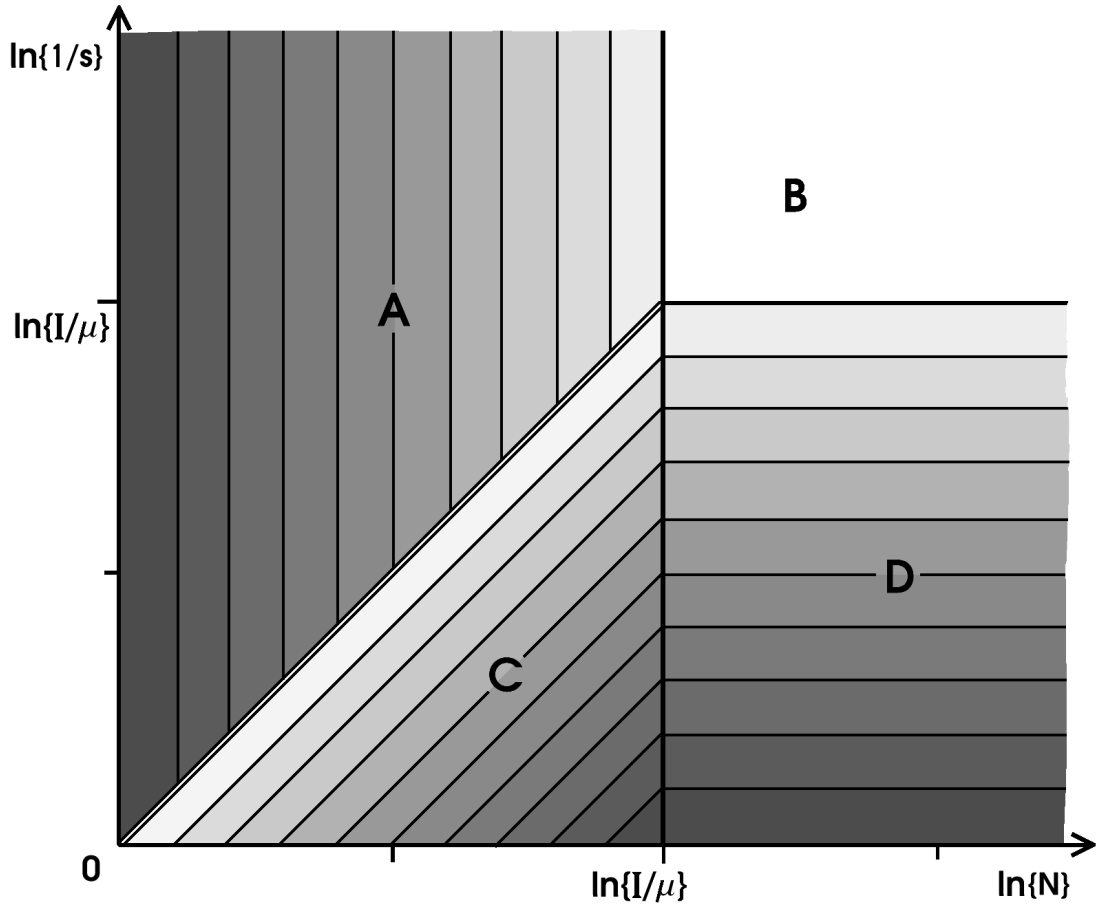, width=15 cm,
}\caption
{Contours for timescales of random fluctuations (in domain A)
and of steady evolution (in domains B,C,D) in a rough
logarithmic plot of the inverse of the Darwinian selection 
coefficient $\ss$ against population number $\NN$, with 
the scale fixed by the inverse of the relevant mutation rate,
$\rr=\mur/\Inf\, ,$ using lighter shading for longer timescales.
{\color{turq} The domains for which the selection pressure is too 
weak to be effective are A, the regime of genetic drift in a small 
population, and B, the regime of genetic equilibrium in a large 
population. The domains of strong selection pressure are C, the
neo-Darwinian regime, with evolution rate limited by rarity of 
mutations in a small population, and D, the classical Darwinian 
regime, in which the population is so large that the required 
mutations are always available.}
\label{FigZones}
}
\end{figure}

In any such system, if the rate $\rr$ of mutation per replication
at a base position of interest is typical, then  knowlege of the 
total mutation rate for the whole genome, namely  
$\mur={\calM} \rr \, ,$ will provide the estimate 
{\be \rr= \mur/\Inf \, . \label{Uu}\fe}
Since, in an initially well adapted species, random mutations
will be seldom favourable and often fatal, the mechanism of
replication must have become reliable enough to ensure that
$\mur$ is not too large compared with unity, while on the other 
hand there is no advantage in taking the trouble to get a much
lower value. It is therefor to be guessed, and seems to be 
observed (Maynard Smith 1989) that under  naturel (as distinct from 
laboratory) conditions $\mur$ will typically have a value
 that is large, but not extremely large, compared with unity. 

In addition to this relatively circumscribed mutation rate 
$\rr \, $ the rate of evolution  at a particular base position will 
be predominantly governed by two other much more widely ranging 
parameters, which are the size $\NN$ of the pertinent interbreeding 
population, and the value of the relevant Darwinian selection 
coefficient $\ss\, ,$ whose possible values are plotted logarithmically
(with repect to the scale determined by the approximately fixed
value of $\rr$) in Figure \ref{FigZones} (which corrects 
the corresponding figure in my earlier work (Carter 1983) where
the -- then less well known -- effect of fluctuations in small 
populations was underestimated).

The relative values of the the three independent parameters,
or the three corresponding numbers, namely
the replication reliability number $\NN_{\!\rr}=1/\rr$
(representing the number of generations over which the base
can be copied without substantial risk of error), the
Darwinian selection number defined as $\NN_{\!\ss}=1/\ss $ 
(indicating how many generations will be needed for selection to 
have a substantial effect), together with the relevant
(interbreeding) population size $\NN\, ,$ will determine four 
qualitatively different regimes, labeled A,B,C,D, in  Figure  
\ref{FigZones}.

The simplest is the classical Darwinian regime, labelled D, 
for which evolution from an ``old'' state to the favoured 
``new'' state at the base position in question will proceed 
with a timescale given roughly in terms of the relevant 
generation timescale $\ttau_{\!g}$ simply by
{\be\ttau_{\!\ss}\approx \NN_{\!\ss}\,\ttau_{\!g}\, .\label{Dar}\fe} 
This applies not just to evolution of the state (what 
geneticiens call an allele) at a single base position
but also to  parallel evolution at many base positions,
provided there is a sufficiently effective mechanism for the
interchange of genetic information, which is of course what is
achieved by sexual reproduction (instead of simple cloning).
In order for  (\ref{Dar}) to hold, it is however necessary, 
not only that the mutation rate should be high enough to avoid 
delay in the provision of the raw material for selection, 
which requires
{\be \NN_{\!\ss} \lta\NN_{\!\rr}\, ,\fe}
but also that the population should be large enough for the effect
of random fluctuations to be negligible, which requires
{\be \NN\gta\NN_{\!\rr}\, , \label{largepop}\fe}
an inequality that is seldom satisfied by large animal populations.
Assuming that $\mur$ is large but not extremely large compared with
unity, the coincidence (\ref{firstcoin}) can be interpreted
as telling us that the human population actually is  marginally large
enough to satisfy this condition now, but it was far too small to do 
so at the time with which we shall be concerned, when our `homo 
sapiens' species first emerged.

The next simplest possibility is that of the mutation controled 
regime,labelled B, for which the large population
condition (\ref{largepop}) is satisfied in conjunction with the 
selective neutrality condition
{\be \NN_{\!\ss} \gta\NN_{\!\rr}\, ,\fe}
which means that Darwinian selection is too weak to be relevant.
In this case replacement of the ``old'' state by the `new'' one will 
proceed with  timescale given directly by the mutation rate as 
{\be\ttau_{\!\rr}\approx \NN_{\!\rr}\,\ttau_{\!g}\, , 
\label{mutime}\fe}
until the attainment of a mixed equilibrium for which the rate of 
reverse mutations balances that of forward mutations.

The situation is more complicated in the fluctuation dominated
neutral regime labelled A, of which the importance was pointed 
out by Kimura.  This regime is characterised by the weak selection 
condition
{\be \NN_{\!\ss}\gta\NN\, ,\fe}
in conjunction with the small population condition (\ref{smallpop}). 
In such circumstances, for the usual (Fisher-Wright) kind of random 
breeding model, the number $n$ of individuals characterised by the 
``new'' state will follow a random walk with step size equal to 
$\sqrt{n}$. Starting from a single ``new'' type individual created 
by a mutation, the probability that such a walk will lead to a 
complete take over, $n\simeq \NN\, ,$  of the ``new'' type 
(instead of leading to its extinction) is only $\approx 1/\NN$, 
and if that occurs this ``fixing'' process will take place with a 
fluctuation timescale 
{\be \ttau_f \approx \NN \ttau_{\! g}\, .\fe}
Since the total rate per generation of occurrence of new mutations in 
the population will be $\NN\rr\, ,$ and the chances of ``fixing'' are 
of order $1/\NN$  -- a factor overlooked in my earlier discussion 
(Carter 1983) -- it follows that the rate per generation of 
establishment of new mutations will  simply be $\rr$. This means that 
a timescale $\ttau_{\!\rr}$ of the relatively long value (\ref{mutime}) 
will separate random events  whereby a fluctuation on the shorter 
timescale $\ttau_f$ will take the entire population from the ``old'' 
to the ``new'' state or back again.

The remaining possibility is  the regime  labelled C, which is  that 
of what Gillespie (2004) refers it as ``positive selection''
in a small population. This regime (the one  most relevant for our 
present purpose) is characterised by the strong selection condition
 {\be \NN_{\!\ss}\lta\NN\, ,\fe}
in conjunction with the small population condition (\ref{smallpop})
{\color{turq} (which characterised the human population until very 
recently)}. In such circumstances, starting from a single ``new'' 
type individual, the number $n$ of such individuals will undergo a 
random walk as in the regime A, but in order for the new type to be 
``fixed'' it is not necessary for such a walk to reach a value 
$n\simeq \NN \ ,$ but just to survive longer than the timescale 
$\ttau_\ss$ needed for the selection effect to become substantial. 
That means reaching a value $n\simeq\NN_{\!\ss}\, ,$ which will occur 
with proportionally higher probability, of order $\ss$. Since the 
total rate per generation of occurrence of new mutations in the
population will be $\NN\rr\, ,$ the rate per generation of 
establishment of new mutations will  be  $\NN\rr\ss$.
This means that evolution from the ``old'' state to the favoured 
``new'' state at the base position in question will proceed with 
a timescale given roughly by
{\be \ttau\approx \ttau_{\!\ss}\, \NN_{\!\rr}/\NN \, . 
\label{neoDar}\fe}

As remarked above, the total mutation rate $\mur$ in (\ref{Uu})
$\mur$ can be expected to be large but only 
moderately large compared with unity, while $\ss$ can be expected 
to be at most moderately small compared with unity, so the
condition
{\be \mur\ss\lta 1 \fe} 
can be expected to hold as a  rough upper limit on 
the rate of genetic evolution under natural conditions.
It therefor follows from (\ref{neoDar}) that
evolution at the maximum allowable rate will therefore
be characterised by a timescale $\ttau$ given by the formula
{\be \ttau\approx \ttau_{\!g}\,\Inf/\NN\, .\label{ttaumin}\fe}

The way the foregoing formulae depend on the population size
$\NN$ suggests that instead of using ordinary time $\tt$ it will 
be informative to reformulate them in terms of population time
$\calT$ as  defined by (\ref{poptimerate}).
This means that the ordinary timescale $\ttau$ in 
(\ref{neoDar}) will correspond to a population timescale
$\Tee$ that is defined by the prescription 
{\be \Tee=\NN\ \ttau\, .\fe}
This population timescale for neo-Darwinian evolution will be 
given, as a rough estimate, by
{\be \Tee\approx  \ttau_{\!\ss}\, \NN_{\!\rr} \, ,\fe}
and it will have a naturally attainable minimum given 
according to (\ref{ttaumin}) by 
{\be \Tee\approx \ttau_{\!g}\,\Inf\, .\label{Teemin}\fe}

According to the palaeological evidence (Falk 1998; Holloway 2001)
as presented in Figure \ref{poplot} hominid cranial expansion 
was characterised, before the emergence of our own species, by a 
timescale $ \ttau $ -- of the order of a few hundred thousand 
years -- that actually does roughly satisfy the rapid evolution 
condition  (\ref{ttaumin}), so that (\ref{Teemin}) does indeed 
hold in this case. It still, however, remains unclear just how 
or why this natural limit was attained.

\bigskip
\noindent
{\bf 7. The memetic breakthrough}
\medskip

The Foerster formula (\ref{quadgrowth}) can be interpreted as
telling us that, as an empirically observed fact, the timescale 
{\be \ttau=\NN/\dot\NN \label{Mal}\fe}
characterising human population growth has been given  in order
of magnitude simply by
{\be \ttau\approx \Tee_{\!\star}/\NN \, .\label{memetime}\fe}
It is evident that that the mechanism for this growth has been
the occupation of a progressively wider range of ecological
niches that have been openned up the introduction of new 
memes of a technical nature, such as those involved in the
use of fire, and subsequently  of cloths, followed more recently
by those involved in crop cultivation, from the invention of
the plough to the latest developments of genetically modified seeds.

It is understandable (Koratayev 2005) that the timescale for this
progress should have evolved roughly in
inverse proportion to the number of people involved, because
the rate of innovation  will presumably  be proportional to
the number of inventive people, who will constitute
what can be expected to be a roughly fixed proportion of
the total population so long as the intrinsic genetic nature
of the population remains the same. What has not been previously
explained is why the proportionality coefficient 
$\Tee_{\!\star}$ in (\ref{memetime}) has the particular value 
characterised by the coincidence (\ref{secondcoin}).

A formula of the form (\ref{memetime}) would presumably have been
applicable to memetic progress (such as the archaeologically
observed improvements in stone chipping techniques), even before
such progress was sufficient to enable substantial population
increase according to (\ref{Mal}) (not just the fitness needed
for survival in intraspecific competition), and thus long
before the emergence of our own `homo sapiens' species,
but with a coefficient $\Tee_{\!\star}$ that would not have
been constant but variable, as a function of the changing
capabilities of the earlier hominids involved, of which
the most important example is that of ``homo erectus'',
the predominant species about a million years ago.  During this 
first phase of hominid evolution the palaeologically measurable 
increase in hominid brain size (Falk 1998; Holloway 2001)
at the rate ultimately characterised by (\ref{Teemin}), as shown 
in Figure \ref{poplot}, would presumable have been
accompanied by a corresponding increase in mental capabilities.
This genetic progress would in turn have brought about a
corresponding decrease in the timescale $\Tee_{\!\star}$
in the formula (\ref{memetime}) for the rate of memetic progress.

\bigskip
\noindent
{\bf 8. Effacement of genetic progress by memetic cross-linkage.}
\medskip

It can now be seen that the coincidence (\ref{secondcoin}) simply
tells us that the decrease in $\Tee_{\!\star}$ came to a halt
when this timescale reached the magnitude $\Tee$ given by
(\ref{Teemin}), in other words when the rate of memetic progress 
overtook the rate of genetic progress. It thus transpires that this
memetic breakthrough is the event that marked the origin of
our own species, since when -- according to the palaeological
record as shown by Figure \ref{poplot} -- there has been a halt not 
only in the decrease in $\Tee_{\!\star}$ but also in the development
of other, more directly observable, genetic features
 of which the most directly observable is brain size.

In the human case, as in the breeding of domestic animals, starting 
with that of dogs from wolves, changing proportions of pre-existing 
genes have brought about many minor variations giving rise to what 
might be described as subspecies. Nevertheless, subsequent to the 
emergence just a few hundred thousand years ago of the particular 
hominid clad to which we belong, there do not seem to have been any 
qualitative  developments of a nature so fundamantal as to have 
required many new mutations.  That is what justifies the 
consensus that all the members of this clad should be classified 
as belonging to one and the same species, namely `homo sapiens'.

This remarkable conspecificity calls for comment. Although this clad 
of ours is relatively young in terms of ordinary time $\tt\, ,$ it 
can be seen from  Figure \ref{poplot} that it is already 
relatively old in terms of population time $\calT \, ,$ as a 
consequence of the tremendous population expansion that has 
taken place in rough accordance with  the Foerster formula 
(\ref{quadgrowth}), initially at the expense of other clads 
(notably the Neanderthals) and later just by occupation of 
the increasing range of ecological niches made available by memetic 
technical progress. According to (\ref{18}) and (\ref{tenfactor})
above, the population time $\calT$ that has elapsed since the 
emergence of `homo sapiens' is about ten times larger than the 
population timescale of order $\Tee_{\star}$ that suffices for 
significant brain growth in the immediately preceeding period. The 
absence of substantial genetic progress since then is therefor 
something that requires explanation.
 
To understand the genetically stagnant nature of the  memetically 
dynamic Foerster phase over such a long population time, 
{\be \calT\simeq 10\,\Tee_{\!\star} \, ,\label{tentime}\fe}
it is instructive to reflect on the purpose of sex,
which functions by  breaking the linkage between different 
sites on the genome. Non-sexual reproduction is common in
microbes and possible, not just by laboratory cloning,
but even under natural conditions, for large animals
such as birds, which can occasionaly pass on their genes
 by virgin birth from unfertilised eggs. However that has
the disadvantage of slowing evolution by requiring that
selection of independent mutations at different sites
should be carried out successively, not simultaneously
as in the usual sexual case.

The question of whether selection is successive or 
simultaneous arises also for memes. Some kinds of memes
are tranmissable independently so as to be simultaneously
selectable like genes in the sexual case. However many
other kinds tend to be linked together in packages within which
weakly favoured innovations will be held up pending 
imposition of those that are  more strongly favoured. 
Furthermore, as well as the possibility of memes
being linked to other memes, it is also common for memes
to be cross-linked with genes. For example one's preference, as a 
sporting participant or spectator, for the tennis meme rather than 
the golf meme, is likely to be independent of one's preference,
at table, for the fork meme rather than the chopstick meme. 
However the latter is likely to be strongly linked with one's
genetic inheritance, because (unlike sports, which are taken up
later in life) eating techniques are usually learned
in infancy from one's own parents.

Although  memes (such as the use of stones for breaking nuts or
mollusc shells) are commonly utilised by other mammals and birds,
ours is the first -- and so far the only --  species for which 
memes have become much more important than genes in the competition 
for survival. Since the breakthrough when this stage of memetic
dominance was reached, there will have been
 a strong tendency for selectively favoured memes to have
carried with them, as involuntary ``hitchhikers'', whatever genes
happenned to be present in the subpopulation in which
the favoured memes were first invented. Such accidentally
favoured genes will thus have an unfair advantage over other
genes (such as those for increased brain size) that might
otherwise have been favoured by systematic selection.
Thus the signals indicating the intrinsically preferred 
direction of genetic progress (notably that of greater braininess) 
will have been effaced  by
memetic noise, which brings about random genetic fluctuations
similar to those that would occur anyway for the very small
populations of the regime labelled A in Figure \ref{FigZones}.

It seems reasonable to conclude that such effacement of genetic 
evolution by memetic ``hitchhiking'' noise is what is responsible 
for the evident lack of substantial genetic progress during the long
population time $\calT$ that has elapsed since the memetic 
breakthrough at the vaguely defined ordinary time,  $\tt_{\!_1}$ say,
when our species first emerged. It is for this reason that
evolutionary divergences since the original emergence of 
`homo sapiens' may occasionally have been sufficient for 
qualification of  temporarily isolated descendent populations
(such as that of the Australian continent) for classification 
as subspecies, but never sufficient for their qualification as 
fully distinct species. 

The straightforeward logical explicability  of this well known 
characteristic of our own terrestrial case, and of the concomitant 
constancy of the Foerster coefficient (without recourse to 
ad hoc hypotheses or reference to special circumstances such as 
climate variations) suggests that these features are not peculiar 
to our particular example, but also relevant in an analogous manner, 
mutatis mutandis,  to other conceivable extrasolar life systems.

\bigskip
\noindent
{\bf 9. Conclusions and implications}
\medskip

To account for the extraordinary sequence of events whereby 
civilisable mankind emerged from the mammalian primate family, 
it is common to seek explanations in terms of  particular
circumstances such as climate change in the neighbourhood of the 
African rift valley.
However as such incidental environmental fluctuations
 had long been going on at various locations on the
drifting continents without producing comparable consequences,
it is reasonable to look for the essential reasons elsewhere.
 The intention of present study has been
to concentrate rather on long term tendencies 
(averaged over such incidental fluctuations)
that may be explicable as generic processes that
could be responsible for the rise of analogous
civilisations in extrasolar planetary systems.

This work has made some headway in the elucidation of the way 
genetic evolution -- particularly brain expansion -- was 
superseded by memetic technical progress {\color{turq} --
shortly after attainment of the level needed for significant
expansion of the hominid population} -- when our 
own `species' emerged a few hundred thousand years ago. 
It transpires that this ``memetic breakthrough''
was of a causally predetermined nature, largely independent of 
special circumstances (such as climate variations) with the 
implication that this 
was not one of the hard-steps referred to 
in the introduction, and that other civilisations in extrasolar
planetary systems might have originated in a similar manner.

What remains as a mystery
is the process whereby the preceding phase of very (maximally) 
rapid brain evolution was initiated, a few million years ago. 
{\color{turq} That transition -- marking the origin of the 
`homo' genus -- stands out as a plausible candidate for  
hard-step status, meaning that the a priori odds may have been 
against its occurrence at all in the available time window.}
Whatever its cause, the maximally rapid nature of this 
evolutionary brain expansion process accounts for the observed 
coincidence (\ref{secondcoin}) whereby the  relevant Foerster 
timescale $\Tee_{\!\star}$ can be roughly accounted for as the 
product of the generation timescale $\ttau_{\!g}$ with the genome 
information content $\Inf\, .$

The other coincidence (\ref{firstcoin}), namely
that the population size $\NN$ has reached that order of
magnitude now, therefore means 
simply that we are now the transition stage, characterised by
$\NN\ttau_{\!g}\simeq \Tee_{\!\star}\, .$
This is when the Foerster phase of 
hyperbolic population expansion must end, because the expansion 
timescale $\NN/\dot\NN$ has fallen below its Malthusian minimum, 
namely the generation timescale $\ttau_{\!g}\, .$ 
Since modern life expectation has risen to about three times the
generation timescale  $\ttau_{\!g}$ the total population
time $\calT$ lived by the witnesses of this transition 
can therefor  be estimated as about $3  \Tee_{\!\star}\, .$
This is nearly a third of all the population time $\calT$ lived by 
humans during the 
Foerster phase that has just come to an end, since our estimate
(\ref{tentime}) for the latter is only about $10  \Tee_{\!\star} \, ,$
which accords with the supposition that
our status as members of this class of witnesses is not particularly
exceptional. However, according to the anthropic principle, that
supposition also requires that the future population time
$\calT$ to be lived by humans should be subject to a bound
of the same order of magnitude, about   $10  \Tee_{\!\star} \, .$

\begin{figure}
\centering
\epsfig{figure=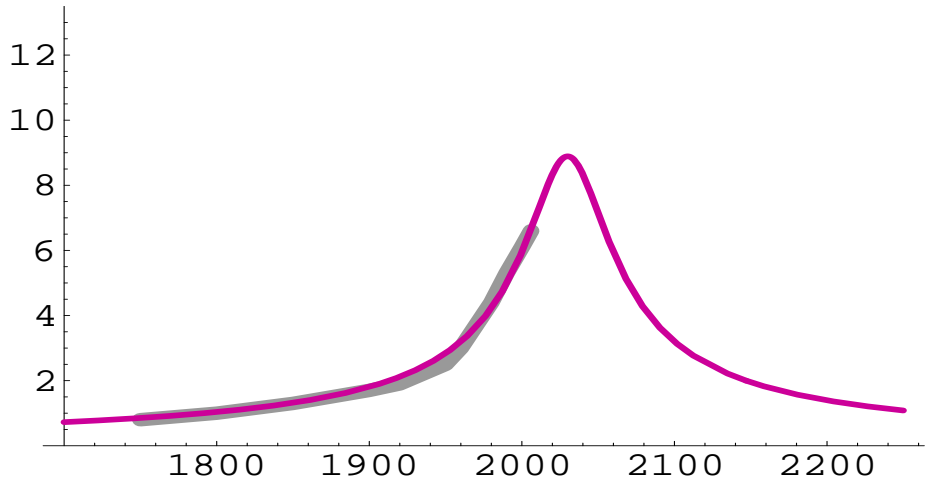, height=6.4 cm, width=12.8 cm}
\caption
{Global population $\NN$, in units of  $10^9$, plotted against date.
Thick pale curve shows U.N. data up to 2000 A.D. Medium curve 
shows how data is well matched by canonical model (\ref{20}), 
with $\Tee_\star\simeq 24\times 10^{10}$ yr, as for Foerster model
in  Figure \ref{FigFoerst}, but with peak about  2035 A.D., and 
with smoothing timescale  $\ttau_{\!g}=27$ yr. The population time 
$\calT$ (area under the curve) will ultimately grow only very 
slowly (as  logarithm of ordinary time $\tt$) so cut-off can be
postponed until distant future.
\label{CanFig}
}
\end{figure}
This anthropic bound -- a few tens of Giga centuries --
on the future population time $\calT$ is automatically 
satisfied by the exponentially decaying Hubbert type extrapolation 
shown in Figure \ref{FigFoerst}, but it is consistent with a 
short-run extrapolation of the more commonly considered  
Verhul{\color{turq}st} type shown in Figure 
\ref{extrapolate} only if its monotonic rise is subsequently
terminated by a ``doomsday'' cut-off of the kind envisaged by 
Leslie (1996) within a century or so at most. The danger 
of mutual extermination by thermonuclear weapons seems lower 
today than it did during the cold war (Shute 1957), but the 
likelihood of a man made ecological catastrophe (Carter 1983)
seems as high as ever. It is however to be emphasized that 
the anthropic bound applies not to the number of individual lives 
or ``souls'', which could be relatively large if average life 
expectation again becomes as short as it was in the past, nor 
-- as emphasized by Wells (2009) in his revision of earlier 
reasoning by Gott (1993) -- to the ordinary time $\tt$ of survival 
of our species, which could be very long if the population size 
becomes sufficiently low.

The admissible time duration of our species could extend as far in 
the future as it did in the past (hundreds of thousands of years) if, 
after passing through a peak in the near future, the population 
undergoes a decline in a manner that mirrors its previous rise. The 
simplest plausible example of such a prolongation is provided, as 
shown in Figure \ref{CanFig}, by a prescription of the canonical form
{\be \NN= {\Tee_\star}/{
\sqrt{(\tt_c-\tt)^2+\ttau_{\!g}^{\,2}}}\, , \label{20}\fe}
in which the time is calibrated, not with respect to a Foerster type 
divergence date $\tt_d\, ,$ but with respect to a moment of time 
symmetry at a new critical date $\tt_c$ when the population passes 
through a smooth peak with width characterised by the generation 
timescale $\ttau_{\!g} \, .$  

The population decline in a scenario of this canonical type
would  have to be rather rapid to begin with, but it would 
soon level out so as to become comfortably imperceptible. Such a 
model might reasonably be chosen as a target of long term public 
policy aimed at coping smoothly with problems of consumption of 
non-renewable resources. However, to fulfill the anthropic 
prediction, if history is a reliable guide, it is more likely that  
monatonic growth of the global population  will go on until it is 
terminated discontinuously by an overshoot catastrophe (Catton 1980) 
of the kind exemplified in the nineteenth century by the Irish 
potato famine.

\vfill\eject

\end{document}